\documentclass[conference]{IEEEtran}
\IEEEoverridecommandlockouts
\usepackage{cite}
\usepackage{amsmath,amssymb,amsfonts}
\usepackage{graphicx}
\usepackage{textcomp}
\usepackage{amsthm}

 \newtheorem*{handshake}{Handshake lemma}
\usepackage{amsfonts}
\usepackage{algorithm}
\usepackage{enumitem}
\usepackage{algpseudocode}
 
 \usepackage{xspace}
\newcommand{\eg}[0]{\textit{e.g.},\xspace}
\newcommand{\ie}[0]{\textit{i.e.},\xspace}
\newcommand{\systemname}[0]{\textsc{Prometheus}\xspace}
\usepackage{xcolor}
\def\BibTeX{{\rm B\kern-.05em{\sc i\kern-.025em b}\kern-.08em
    T\kern-.1667em\lower.7ex\hbox{E}\kern-.125emX}}
\usepackage[hyphens]{url}
 \usepackage{hyperref}

\newcommand{\smartparagraph}[1]{\vspace{.05in}\noindent\textbf{#1}}
    
\begin{document}
\bstctlcite{IEEEexample:BSTcontrol}

\title{From Scientific Texts to Verifiable Code: \\Automating the Process with Transformers
}

 \author{\IEEEauthorblockN{Changjie Wang}
 \IEEEauthorblockA{KTH Royal Institute of Technology}
 \and
 \IEEEauthorblockN{Mariano Scazzariello}
 \IEEEauthorblockA{KTH Royal Institute of Technology}
 \and
 \IEEEauthorblockN{Marco Chiesa}
 \IEEEauthorblockA{KTH Royal Institute of Technology} 
 }

\maketitle

\begin{abstract}
Despite the vast body of research literature proposing algorithms with formal guarantees, the amount of verifiable code in today’s systems remains minimal. This discrepancy stems from the inherent difficulty of verifying code, particularly due to the time-consuming nature and strict formalism of proof details that formal verification tools require. However, the emergence of transformers in Large Language Models presents a promising solution to this challenge. In this position paper, we believe that transformers have the potential to read research papers that propose algorithms with formal proofs and translate these proofs into verifiable code. We leverage transformers to first build a formal structure of the proof using the original text from the paper, and then to handle the tedious, low-level aspects of proofs that are often omitted by humans. We argue that this approach can significantly reduce the barrier to formal verification. The above idea of reading papers to write verifiable code opens new avenues for automating the verification of complex systems, enabling a future where formally verified algorithms from academic research can more seamlessly transition into real-world software systems, thereby improving code reliability and security.
\end{abstract}


\section{Introduction}
Throughout history, humans have continually developed more sophisticated algorithms to solve complex problems, uncovering and forging connections between concepts in innovative and complex ways, and used this knowledge to build the software that runs in modern-day systems. The culmination of all this research knowledge is captured in human language and recorded in countless papers, books, or other written works. Despite this vast body of algorithmic knowledge, only a negligible fraction of today's code in systems has been formally verified using computers~\cite{formal_veri_not_adopt}. The reasons for this lie in the inherent complexity and challenges involved in the formal verification process~\cite{verification_complexity}.

\smartparagraph{Formal verification is time-consuming and complicated. } This complexity arises from several factors. First, it requires familiarity with specialized formal languages, which have their own syntax and semantics that require an enormous amount of time to master adequately~\cite{formal_veri_hard_to_learn}. Second, the formal verification process operates at an extremely low level; every step, even those that are intuitively obvious to humans, must be explicitly verified — a tedious and often frustrating task~\cite{proof_assistant_difficult}. Third, verifiers attempt to aid this process by using heuristics to link logical relationships, but the reliance on heuristics introduces something known as \mbox{\textit{brittleness}}~\cite{brittleness}. Even a minor change, like renaming a variable or removing seemingly irrelevant assertions, can cause previously verified proofs to fail. Finally, even if one could skip proving low-level lemmas, translating proofs from human language into high-level verifiable code remains time-consuming. This is because proofs may be vague and long, covering several pages of an article or book. The complexity of the structure of the proof and properties of data types quickly escalates, making it difficult for humans to grasp the complexity. Existing solutions that try to automate formal verification have significant limitations, such as the inability to read text and scale to simple proofs or code bases~\cite{proof_assistant_natural_langauge}. 

\smartparagraph{Transformers to the rescue?} Transformers are a type of neural network architecture designed to \textit{quickly} process and generate sequences (\eg text)~\cite{attention-is-all-u-need}, excelling at understanding and manipulating human language, which makes them well-suited for tasks like translating textual descriptions into code~\cite{llm4code}.
%
Tools like OpenAI's Codex~\cite{codex} and GitHub Copilot~\cite{copilot} leverage these models to suggest code — with some levels of accuracy — but they do \textit{not} perform formal verification of the produced code, leaving correctness not guaranteed.
As our experiments further reveal, naively applying a transformer-based LLM to the formal verification problem fails to produce any verifiable proofs, due to the low-level complexities. Furthermore, even existing transformer-based systems for writing verifiable code barely scale to simple programs that, for example, contain a single loop~\cite{sun2024clover,misu2024towards,mugnier2024laurel}.

\smartparagraph{Our vision: from textual descriptions to verifiable code using transformers and separation of responsibilities.}
We argue that transformers are still a natural candidate for tackling the challenge of converting algorithms and proofs from human language into verification code. We, however, need to use them for the right task. Our key intuition is that we can break down the verification task into two main stages:
\begin{enumerate}[leftmargin=*,noitemsep]
\item \textit{Translate a textual proof from human language to high-level verification code.} 
Transformers excel at translations that are roughly one-to-one (\eg between spoken languages), making them well-suited for converting a textual proof into a high-level skeleton of code for the verifier. At this stage, the goal is not to complete the entire proof but to outline its structure, with many low-level details intentionally left out. The key insight here is that verifiers can prove that the high-level proof works assuming (without proving) that the details of the low-level proofs are correct. This allows us to use the best aspect of transformers for building the skeleton of the formal proof while deferring the detailed, low-level proofs to a second stage.
\item \textit{Generate the missing low-level verifiable proofs.}  For the second stage, we claim that transformers are still the right tool. We use them to autonomously generate low-level proofs that lack textual equivalents, typically because their solutions are intuitively obvious to humans. Our key observation is that such low-level proofs do not require deep intellectual effort. Were they more complex, they would need explicit descriptions in text to ensure understanding from human readers. This observation allows us to leverage the strengths of current transformers, which, while limited in complex reasoning~\cite{llm_lack_reasoning}, are perfectly suited for this type of task. Additionally, since many of these proofs exhibit recurring patterns, one can envision training or fine-tuning specialized, smaller transformers on these specific types of proofs. This pre-training enables the transformer to efficiently produce these necessary but straightforward verifications, streamlining and complementing the high-level formal verification from the previous stage.
\end{enumerate}

\smartparagraph{Challenges and a feasibility prototype.} Beyond laying out a vision, this paper explores the challenges associated with generating formal proof of correctness for both experts and transformer-based LLMs. We discuss some of these challenges through selective examples that are relevant for network system verification, whereas a comprehensive list would be beyond the scope of this work. We provide hints that a transformed-based LLM would be a suitable tool for producing a complete formal proof, limiting human intervention to only verifying the correctness of the formal specifications (not the proofs). 
To support our position paper, we show how to verify three selected network topological properties using a simple LLM-based prototype called \systemname.\footnote{Prometheus is a figure in Greek mythology who stole ``fire'' from the gods and gave it to humanity in the form of technology and knowledge.} Our prototype primarily focuses on network properties derived from graph theory.
 \footnote{Our insights would apply to potentially different domains.} 
Specifically, through our analysis, we highlight how even a seemingly straightforward lemma, such as the ``handshake lemma'' from graph theory, would require substantial verification efforts for those not specialized in the field, but can be proved using our prototype. Our results show that \systemname outperforms existing LLM-based verifiers thanks to its ability in separating the proof verification into a high-level and low-level verification synthesis process. We focus on code verification and tools more widely used by developers (\textit{e.g.}, Dafny), thus we do not discuss related work on the pure math verification or other verifiers (\textit{e.g.},~\cite{proofsketch}). 

\smartparagraph{Impact.} We believe our study has several far-reaching implications. First, it opens the possibility of automatically verifying the correctness of decades of academic research, an ambitious undertaking in itself. Second, by integrating our system, researchers will have the ability to formalize proofs in an easier way, with an automated transformer providing verification. This process would enhance transparency, allowing both authors and reviewers to verify the correctness of claims more effectively. Third, based on the extensive formal knowledge generated by our approach, we can envision developing systems that not only verify but also create \textit{novel} algorithms with proofs.

\section{Example: The handshake lemma}\label{sect:handshake-lemma}
We explain the main challenges in writing verification code using a simple network topological property that we derive from graph theory~\cite{book-graph-theory}:

\begin{handshake}\label{lem:handshake}For every graph $G = (V, E)$ we have $2|E| = \sum_{v \in V} d(v)$, where $d(v)$ is the degree of vertex $v$.
\end{handshake}

This lemma states that the sum of the degrees of all vertices is equal to twice the number of edges in the graph. Even without looking at the proof, a human immediately guesses the correctness of the statement with the following reasoning:

\vspace{-0.05in}
\begin{proof}[Human intuition]Each edge connects \textit{two} vertices, increasing  the overall sum of the degrees of all the graph vertices by two.
\renewcommand{\qedsymbol}{}
\end{proof}

\vspace{-0.2in}
This reasoning is generally enough to convince the reader about the correctness of the lemma. Some books may provide some additional proof text as in the following example:

\vspace{-0.05in}
\begin{proof}[Book proof~\cite{book-graph-theory}]Let $X = \{(e, x) : e \in E, x \in V, x \in e\}$.  Then
$|X| = \sum_{v\in V} d(v)$ and $|X| = \sum_{e\in E} 2 = 2|E|$.
\end{proof}

\vspace{-0.05in}
This proof more formally defines human intuition using an auxiliary set $X$, which enumerates all the edge-vertex pairs of the graph. It then claims that the cardinality of this set (\ie $X$) is equal to both (i)  the sum of the degrees of the vertices, and (ii) the sum of the edges times two.

In this example, for the sake of simplicity, we only focus on generating verification code (\textit{i.e.}, the proof), rather than the code of an algorithm. But, the principles behind \systemname could be extended to enable joint generation of both algorithms and their corresponding proofs.  

\smartparagraph{Why is it so hard to verify the lemma for a non-expert?} Neither the human intuition nor the above book proof can be easily \textit{formally} verified. Consider the more formal proof from the book. Even if we write the complete formal specification of the problem, including the property to be proved and a specification of a graph data structure, a verifier like Dafny~\cite{dafny} will not be able to verify these two assertions  $|X| = \sum_{v\in V} d(x)$ and  $|X| = 2|E|$. 
Indeed, formal verifiers operate \textit{without understanding the semantics} or meaning of the mathematical problem that they analyze. Instead, they apply a series of pre-defined heuristics and logical techniques to check whether a given post-condition or specification holds. These tools function purely at the syntactic level, using rules and strategies to attempt proofs, but without any intrinsic comprehension of the underlying, potentially trivial, mathematical concepts or their purpose.
We showcase these limitations from formal verifiers using the $|X(G)|=\sum_{v\in V}d(v)$ assertion from the handshake lemma and use the Dafny verification system\footnote{We focus on the Dafny verifier as it is widely used in the industry~\cite{aws}.}:
\begin{enumerate}
    \item \textit{Definitions are side-effect-free.} Any definition of a property expressed in Dafny must be described \textit{without} using side-effect operations~\cite{dafny-reference-manual}, \ie the definition should not modify the state of the involved variables in the definition. Practically speaking, this constraint forbids a programmer from using ``for'' loops in the definition of properties. As an example, $\sum_{v\in V}d(v)$ cannot be defined with a for loop. Dafny only allows programmers to use recursive definitions to define properties and the recursive function cannot modify any state, only return values. The sum of the degrees of the vertices $S_d$ would be defined as:
    
\vspace{-.16in}
\[ 
S_d(G) = 
\begin{cases} 
    0 & \text{if } V = \emptyset, \\
    d(v) + S_d( (V \setminus \{v\}), E) & \text{else }
\end{cases}
\]
    
    It is key to understand that verifiers reason in this ``recursive space'' and all proofs must align with these definitions. As developers rarely rely on recursion to develop code, this constraint represents a barrier for most developers wishing to formally verify their code.

    \item \textit{Inductions are pervasive: the summation example. } Since most definitions are recursive, it is often required to prove even simple properties using inductive proofs. Let's consider the same example as above, \ie $|X|=\sum_{v\in V}d(v) = S_d(G)$. To prove this property, we would write a proof by induction on the set of vertices.
    In the inductive step, one needs to prove that adding a vertex increases both the sum of degrees and $|X|$ by $d(v)$. While obvious, this inductive step is non-trivial to prove in a verifier as we discuss in the next step. 

    \item \textit{Inductions are pervasive: the set augmentation example.} When we add a vertex in the above inductive step, its degree is defined as the sum of its adjacent edges, \ie $d(v)=|N(v)|$, where $N(v)$ represents the neighbors of $v$ and is defined as $\{e|e\in E, v\in e\}$. 
    At the same time, by the definition of $X$, vertex $v$ contributes with elements $X(v)=\{(e,x)|e\in N(v),x=v\}$. Clearly, the two sets have the same number of elements, \ie $|d(v)| == |X(v)|$ since both sets enumerate the edges adjacent at $v$, but $X(v)$ stores the edges as pairs always repeating $v$ as the second element in the pair. For example, consider a vertex $v$ adjacent to edges $e_1$ and $e_2$. We have that $d(v) = |\{e_1,e_2\}|$ while $|X(v)|=|\{(e_1,v),(e_2,v)\}|$.  A verifier like Dafny does not easily understand that the cardinalities of the two sets are equivalent by the above definitions. It requires another proof by induction over the sets of edges in $N(v)$ and $X(v)$, showing that adding an edge preserves $|N(v)|=|X(v)|$. Even if we write this induction, a human still needs to address another low-level challenge, which we discuss next. 

    \item \textit{Extension equalities.} Following our example, we now must prove $|N(v)|=|X(v)|$ by induction, \ie we prove $|N(v)|=|\{(e,x)|e\in N(v),x=v\}|$. We do not cover the base case. In the inductive step, we remove any edge $k$ from $N(v)$. By inductive hypothesis we have $|N(v)-\{k\}|=|\{(e,x)|e\in (N(v)-\{k\}),x=v\}|$. Dafny easily understands that removing an edge from $N(v)$ results in reducing its cardinality by one. However, Dafny cannot automatically understand the same for $X(v)$. Asking Dafny to prove that  $|\{(e,x)|e\in (N(v)-\{k\}),x=v\} + \{k,v\}| == |\{(e,x)|e\in N(v),x=v\}|$ will fail. The reason is subtle. Dafny relies on some heuristics to find out how to prove assertions. In this case, we are asking a question about the sizes of the sets that are actually the same but expressed differently. We need to help Dafny in proving this assertion by first asking to check if the sets are identical, \ie we must first prove $\{(e,x)|e\in (N(v)-\{k\}),x=v\} + \{k,v\} == \{(e,x)|e\in N(v),x=v\}$ so that the Dafny verifier uses its heuristics to prove this assertion and only then, it can prove the cardinality one. This example illustrates the importance of allowing humans to add intermediate assertions to guide the solver and its heuristics toward correct lemma verification. Here, these guiding assertions relate to \textit{extension equalities} and assist Dafny in recognizing that two objects are, in fact, identical.

\end{enumerate}

\vspace{-.03in}
\smartparagraph{Would an LLM help in proving the handshake lemma?} 
We answer this question by feeding the lemma statement and its textual proof from the book~\cite{book-graph-theory} to the Claude Sonnet 3.5 LLM and asking it to produce Dafny code. Most of the time, Claude will produce the following code:

\vspace{-.13in}
\noindent\rule{\linewidth}{0.4pt}
\begin{algorithmic}[1]
\small
\State \textbf{lemma} \Call{HandshakeLemma}{g: Graph}
\State \textbf{requires} \Call{IsValidGraph}{g}
\State \textbf{ensures} \Call{SumOfDegrees}{g} $== 2\cdot \mid$E$\mid$
\Statex \{
\State \hspace{.5em} var V, E := g.V, g.E;
\State \hspace{.5em} var X := set e, v $\mid$
e in E \& v in V \& \Call{Adj}{e,v} :: (e, v);
\State \hspace{.5em} \textbf{assert} $\mid$X$\mid == $ \Call{SumOfDegrees}{g};
\State \hspace{.5em} \textbf{assert} $\mid$X$\mid == 2 \cdot \mid$E$\mid$;

\Statex \}
\end{algorithmic}
\vspace{-.13in}
\noindent\rule{\linewidth}{0.4pt}

 The translation of the proof is accurate as it covers every logical step, \ie the definition of set $X$ (where \textsc{Adj} is an adjacency predicate) and the two assertions claiming $\lvert X \rvert$ is equal to both the sum of degrees and twice the number of edges. Dafny cannot independently verify these two assertions. Yet, one main contribution of an LLM is that it can lay out the outline of a proof so that one only needs to focus on the low-level technical parts. LLMs are extremely efficient in this type of translation from natural language to Dafny code as they are almost one-to-one, a perfect task for transformers. 

 What it remains to do is writing the low-level proofs showing that the two assertions hold. This is where an LLM needs to find its own ways of proving sub lemmas. We argue that these low-level lemmas do not require complex intellectual reasoning, which would otherwise be expressed in human language for the reader to trust the proof. Instead, these low-level lemmas often require dealing with the low-level time-consuming details of the Dafny language and how data structures are expressed. We show in the next section how a simple LLM-based prototype manages to write Dafny code to also verify the low-level aspects of a proof.


\section{Feasibility Study with a Prototype}
%
We perform a preliminary evaluation of our idea to assess the potential feasibility of the idea, leaving a full-scale evaluation as future work. 
Our prototype feeds a proof in natural language as input to the Claude Sonnet 3.5 LLM. 
It then interacts iteratively with the LLM by proving lemmas using a top-down approach, \ie first prove the general lemma using some lower-level helper lemmas that will be proved in the next iterations. The system also verifies the code generated by the LLM, and sends both the output of the verification process alongside some hints on how to solve these problems. These hints are high-level hints to help the LLM in solving common Dafny problems, \ie proving using inductions, solving extension equality problems, spotting missing pre-conditions, and more. 
We test \systemname on three lemmas:
\begin{itemize}[leftmargin=*,noitemsep]
    \item \textit{Handshake lemma} (definition in \S\ref{sect:handshake-lemma}).
    \item \textit{Degree bounds lemma}. Given a graph $G$, we have that $min(G)\le avg (G) \le max(G)$, where $min(G)$, $avg(G)$, and $max(G)$ are the minimum, average, and maximum degree of the vertices in $G$.
    \item \textit{Even cycle bipartite lemma}. Given a bipartite graph $G=((A,B),E)$, we have that every cycle has an even length, where $A$ and $B$ are two partitions of the vertices such that there are no edges connecting two vertices in the same partition.
\end{itemize}


For each lemma, we input all the formal specifications of all the relevant definitions, such as the graph data structure or the definition of a cycle. To evaluate the performance of \systemname, we use zero-shot prompting as a baseline, requesting that LLMs generate the complete proof in a single attempt. We compare the success rate of \systemname to the baseline as the number of successful runs over the total runs. The following table presents the results for the three lemmas, including the success rate over five runs:

 \vspace{-0.05in}
\begin{table}[h]

\label{tab_success_rate}
\centering
\begin{tabular}{cccc}
\multicolumn{1}{c}{}   & \multicolumn{1}{c}{\textbf{Handshake}} & \multicolumn{1}{c}{\textbf{Degree Bounds}} & \multicolumn{1}{c}{\textbf{Bipartite}} \\ 
\hline
\multicolumn{1}{c}{Baseline}   & \multicolumn{1}{c}{0/5}       & \multicolumn{1}{c}{0/5}          & \multicolumn{1}{c}{0/5} \\
\multicolumn{1}{c}{\textbf{\systemname}} & \multicolumn{1}{c}{\textbf{4/5}}       & \multicolumn{1}{c}{\textbf{3/5}}          & \multicolumn{1}{c}{\textbf{3/5}} \\
\end{tabular}

\end{table}

\vspace{-0.05in}
We observe that with basic prompting strategies like zero-shot prompting, the baseline fails to produce verifiable code. The generated code is always short and lacks low-level proofs mentioned in \S\ref{sect:handshake-lemma}. In contrast, \systemname successfully generates a complete and verifiable proof in Dafny, with a high success rate. \systemname always starts with a proof skeleton, delves into the low-level details, and refines them iteratively. View a video of a successful run~\cite{video-prometheus}.
 The result shows that \systemname delivers a promising performance in verifiable code generation in Dafny.

\section{Future Challenges and Discussion}

\vspace{-0,05in}
\smartparagraph{Who verifies the correctness of the input formal specifications?} In our evaluation, all formal specifications—including data structure and predicate definitions (e.g., how a graph or the sum of node degrees is defined) as well as pre- and post-conditions of the proven property—were generated using an LLM. However, we manually verified the correctness of these definitions before using them as input for \systemname. Completely automating this task is challenging, as it forms the foundational truth that guides the LLM in generating accurate code. Further research on how to more reliably generate formal specifications from natural language is needed. 

\smartparagraph{How far can we go in proving theorems without a textual proof?} 
As we discussed in this paper, there is always a mismatch between the level of detail required by a formal verifier and a textual proof. In \systemname, we try to fill this gap in two ways: either having an LLM generating the missing logic or letting verifiers generate a proof using heuristics. In both cases, neither LLMs nor verifiers possess the same level of ``intelligence'' that humans possess in order to understand relations between concepts, \eg impact of an edge on the sum of degrees of the vertices. Exploring novel ways to generate proofs (as in the AlphaProof or Alpha Geometry 2 systems) using transformers and reinforcement learning is left outside the scope of this paper, which only deals with translating the existing human knowledge into verifiable code. 

\section*{Conclusions}
\vspace{-0.05in}
In this position paper, we argue that recent advancements in LLMs present unprecedented opportunities for translating decades of scientific research on algorithms and systems into formally verifiable code. We propose leveraging LLMs to generate high-level code outlines and corresponding verification logic from natural language descriptions in the literature. This approach aims to lower the barrier to verifying scientific research and enable the creation of more resilient software systems. Significant research challenges remain, including developing systems capable of accurately translating complex notations, algorithms, and logic, as well as innovating stronger methods for verifying the low-level aspects of these proofs.

\section*{Acknowledgments}

We would like to thank the anonymous reviewers for their insightful comments and suggestions on this paper. This work has been partially supported by Vinnova (the Sweden's Innovation Agency), the Swedish Research Council (agreement No. 2021-04212), and KTH Digital Futures.

\sloppy
\bibliographystyle{IEEEtran}
\begin{small}
\bibliography{bibliography}
\end{small}

\end{document}